\documentclass[aps,preprint,showpacs,amsmath,floatfix]{revtex4}
\usepackage{graphicx}%
\usepackage{dcolumn}
\usepackage{color}
\usepackage{amsmath}
\usepackage{here}

\begin{document}
\newcommand{\be}{\begin{equation}}
\newcommand{\ee}{\end{equation}}
\newcommand{\rojo}[1]{\textcolor{red}{#1}}

\title{Nonlinear Surface Impurity in a Semi-infinite Lattice}

\author{M. I.  Molina}

\affiliation{Departamento de F\'{\i}sica, Facultad de
Ciencias, Universidad de Chile, Casilla 653, Santiago, Chile}

\begin{abstract}
We examine the formation of bound states on a generalized nonlinear
impurity located at or near the beginning (surface) of a linear,
tight-binding semi-infinite lattice. Using the formalism of
lattice Green functions, we obtain in closed form the number of
bound states as well as their energies and probability profiles,
for different nonlinearity parameter values and nonlinearity
exponents,  at different distances from the surface. It is shown that
close to the surface, the amount of nonlinearity needed
to create a bound state or to effect dynamical selftrapping,
increases (decreases) depending on whether the exponent is smaller
(larger) than, approximately, two.

\end{abstract}


\maketitle

The interplay of nonlinearity and discreteness has received
considerable attention recently\cite{kv_pt}, since it plays a vital
role in the emergence of a new kind of
excitation in extended, nonlinear systems with discrete
translational invariance, known as intrinsic localized
modes (ILM). These ILMs are generic to
physical systems of interest, such as arrays of nonlinear optical
waveguides\cite{optics}, molecular crystals\cite{crystals},
biopolymers\cite{biopolymers}, arrays of
Josephson junctions\cite{jj} and even Bose-Einstein condensates in
magneto-optical traps\cite{BEC}.

Given the strictly local manner in which nonlinearity enters into the
effective evolution equations in all these cases (see below), one is
led to the idea that in the limit of strong nonlinearity,  one could
approximate a typical nonlinear system by a linear one containing
a small cluster of nonlinear sites, or even a single nonlinear
impurity. The system thus simplified is amenable to exact
mathematical treatment, and the influence of other, potentially
competing effects such as dimensionality, boundary effects, noise,
etc., can be more easily studied without losing the essential
physics.

For a one-dimensional discrete system 
in the presence of a single nonlinear impurity, located at $n = d$, 
the dynamics is given by the well-known discrete nonlinear
Schr\"{o}dinger (DNLS) equation:
\be
i {d C_{n}\over{d t}} = V (
C_{n+1} + C_{n-1}) - \chi |C_{n}|^{\beta} C_{n} \delta_{n,d},
    \hspace{0.5cm}(\hbar \equiv 1)\label{eq:1}
\ee
where $C_{n}$ is the probability amplitude for finding the
excitation on site $n$, $V$ is the nearest-neighbor transfer matrix
element, $\chi$ is the nonlinearity parameter and $\beta$ is the
nonlinearity exponent. Usually, but not always, $\beta = 2$ which in a
condensed mater context corresponds to an underlying harmonic oscillator
degree of freedom `enslaved' to the excitation (electron) at the impurity site.
When this vibrational impurity is anharmonic in nature, other $\beta$ values
are possible in principle, with $\beta < 2$ corresponding to a
``hard'' vibrational impurity while $\beta > 2$ corresponds to a
``soft'' case\cite{vib}.

Bound states for single nonlinear impurities embedded in infinite lattices
include chains\cite{mt_prb,mth_pre,saopaulo}, Cayley trees\cite{cayley},
triangular\cite{busta} and a cubic\cite{hui, busta} lattices.
Now, since the creation of a bound
state, or the dynamical selftraping at the impurity site implies
the localization of energy on a scale of the order of the lattice
spacing,  one might surmise that, by placing the nonlinear
impurity at or near the surface of a semi-infinite lattice, the
nonlinearity strength needed to effect localization would
decrease, facilitating in this way its creation and experimental
observation. As an step in that direction, in this work we examine
a simple model consisting of an electron (or excitation) propagating in a semi-infinite, linear
chain, that contains a single nonlinear impurity at a distance $d$
from the beginning (`surface') of a semi-infinite chain (Fig. 1),
and examine the conditions for the existence
of bound state(s) and the dynamical selftrapping properties, and
compare them to the results obtained for the infinite chain
\cite{mth_pre}. 
\begin{figure}
\centerline{\includegraphics[width=3.5in]{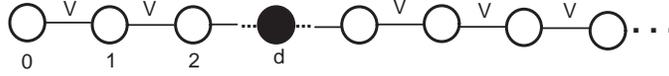}} \caption{A
nonlinear impurity near the surface of a one dimensional chain}
\label{fig1}
\end{figure}

{\em Bound States.}\ We consider Eq.(\ref{eq:1}) for a semi-infinite lattice ($n = 0,1,\cdots$)
and normalize all energies to the half bandwidth of the infinite chain case.
The Hamiltonian is given by  
\be
H = {1\over{2}} \sum_{n=0}^{\infty} ( |n\rangle \langle n+1| + |n+1\rangle \langle n| ) +
    \gamma |C_{d}|^{\beta} |d\rangle \langle d|
\ee where $\{|n\rangle\}$ are Wannier states and $\gamma \equiv
\chi/(2 V)$. The dimensionless Green function $G = 1/(z - H)$ can
be formally expanded as\cite{economou} $G = G^{(0)} + G^{(0)}
H_{1} G^{(0)} + G^{(0)} H_{1} G^{(0)} H_{1} G^{(0)} +
\cdots$, where $G^{(0)}$ is the unperturbed ($\gamma
= 0$) Green function and $H_{1} = \gamma |C_{d}|^{\beta} |d\rangle
\langle d|$. The series can be resumed to all orders to
yield \be G_{m n} = G_{m n}^{(0)} + {\gamma |C_{d}|^{\beta} G_{m
d}^{(0)} G_{d n}^{(0)}\over{1 - \gamma |C_{d}|^{\beta} G_{d
d}^{(0)}}}.\label{eq:Gmn} \ee where $G_{m n} \equiv \langle
m|G|n\rangle$. Now, we cannot use Eq.(\ref{eq:Gmn}) directly since
we do not know $C_{d}$, but we will determine it through an exact
selfconsistent procedure: The energy of the bound state(s) is
obtained form the poles of $G_{m n}$, i.e., by solving $1 = \gamma
|C_{d}|^{\beta} G_{d d}^{(0)}(z_{b})$. On the other hand, the
bound state amplitudes $C_{n}$ are obtained form the residues of
$G_{m n}$ at $z = z_{b}$. In particular, at the impurity site,
$ |C_{d}|^{2} = Res\{ G_{d d}(z)\}_{z = z_{b}} = - {G_{d
d}^{(0)}}^{2}(z_{b})/G_{d d}^{'(0)} (z_{b})$ Inserting
this back into the bound state energy equation leads to \be
{1\over{\gamma}} = {{G_{d d}^{(0)}}^{\beta + 1}(z_{b})\over{[-G_{d
d}^{'(0)} (z_{b})]^{\beta/2}}}.\label{eq:zb} \ee The unperturbed
Green function $G_{m n}^{(0)}$ for the semi-infinite lattice, can
be calculated by the method of mirror images: Since there is no
lattice to the left of $n = 0$, $G_{m n}^{(0)}$ should vanish
identically at $n = -1$. Thus, $G_{m n}^{(0)}(z) = G_{m
n}^{\infty}(z) - G_{m, -n-2}^{\infty}(z)$, where $G_{m
n}^{\infty}(z)$ is the Green function for the infinite lattice,
$G_{m n}^{\infty}(z) = \mbox{sgn}(z)(1/\sqrt{z^{2} -1}) [ z -
\mbox{sgn}(z)\sqrt{z^{2} - 1} ]^{|n-m|}$, where $\mbox{sgn}(z) =
+1(-1)$ for $z>0 (<0)$. Therefore,
\begin{eqnarray}
G_{m n}^{(0)}(z)& = & \mbox{sgn}(z){1\over{\sqrt{z^2 -1}}} [z -
\mbox{sgn}(z)\sqrt{z^2 - 1}]^{|n-m|} \nonumber\\
                &    &
-\mbox{sgn}(z){1\over{\sqrt{z^2 -1}}} [z - \mbox{sgn}(z)\sqrt{z^2
- 1}]^{|n+2+m|}.\label{eq:G0}\ \ \ \ \
\end{eqnarray} 
From Eq.(\ref{eq:G0}) we note the parity property
$G_{dd}^{(0)}(-z) = - G_{dd}^{(0)}(-z)$, which implies
$G_{dd}^{'(0)}(-z) = G_{dd}^{'(0)}(-z)$. This means, according
to Eq.(\ref{eq:zb}) that the change $\gamma \rightarrow -\gamma$,
reverses the sign of $z_{b}$. On the other hand, from
Eq.(\ref{eq:1}), it is possible to deduce that the change $\gamma
\rightarrow -\gamma$ is equivalent to the change $C_{n}\rightarrow
(-1)^{n}\ C_{n}^{*}$. Since we are interested in a localized
state, where the $C_{n}$ can be chosen as real, we conclude that a
change in sign of the nonlinearity parameter reverses both the
``staggered'' character of the bound state and the sign of the
localized state energy.

After inserting
Eq.(\ref{eq:G0}) into (\ref{eq:zb}), the general structure for the
number of bound states emerges: For any finite distance $d$ from
the surface and any positive value of the exponent $\beta$, there
is a critical amount of nonlinearity $\gamma$ below which there is
no bound state, and above which there are two bound states. For
$\beta $ exponents smaller than $2$, and as $d$ is increased, one
of the bound states tends to merge with the band edge, so that in
the limit of a very deep impurity, there is only a single
localized bound state. For $\beta
>2$, however, as $d$ is increased, both bound states remain
localized. In the special case of a linear impurity ($\beta = 0$),
there is a single bound state provided $\gamma > 1/(d + 1)$. Thus,
in the limit $d\rightarrow \infty$ these results are consistent
with the case of a completely infinite lattice\cite{mth_pre}: A
single bound state for $\beta < 2$, and for $\beta > 2$, a critical curve in
nonlinearity strength-nonlinearity exponent space, separating a
region with no bound states from a region with two bound states.
At the surface ($d = 0$) the critical curve is given by \be
\gamma_{c} = {(1 + \beta)^{(1+\beta)/2}\over{2 \beta^{\beta/2}}}
\ee In particular, for the DNLS case, $\gamma_{c} = 3^{3/2}/4
\approx 1.3$, larger than the value for the infinite chain
($\gamma_{c}=1$). Figure 2 shows phase diagrams in $\gamma$-$\beta$
space showing the number of bound states, at different distances
$d$ between the impurity and the `surface' of the system.
\begin{figure}[ht]
\noindent\includegraphics[height=2.7in,width=3.5in]{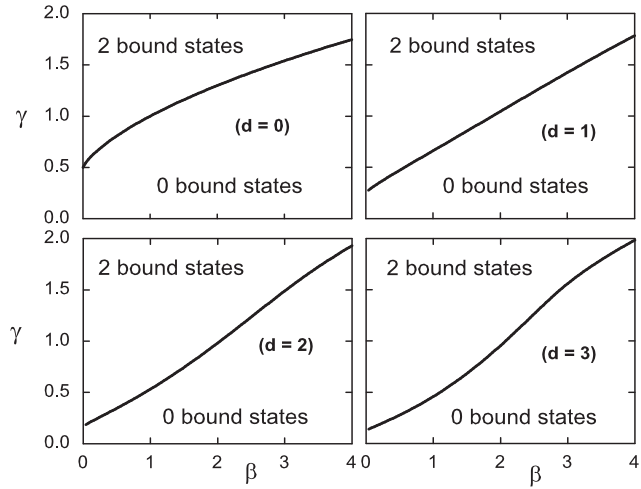}
\caption[]{Phase diagram in $\gamma$-$\beta$ space, showing the
number of bound states, for different distances impurity-surface
(in units of the lattice constant).} \label{fig2}
\end{figure}
\begin{figure}
\noindent\includegraphics[height=2.in,width=3.0in]{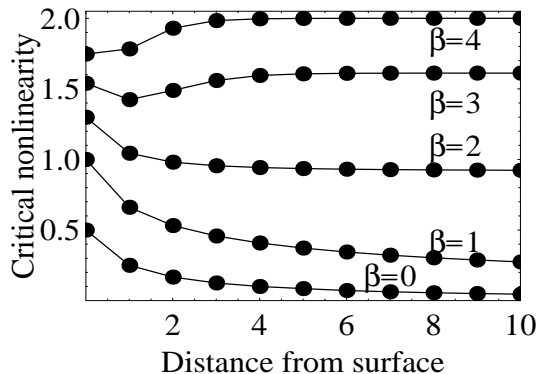}
\caption{Scaled critical nonlinearity for onset of a bound state, as a
function of the distance from the nonlinear impurity to the
`surface' of the chain.} \label{fig3}
\end{figure}

As to the stability of these bound states, it is easy to see from
a graphical analysis of the structure of Eq.(\ref{eq:zb}) that,
as nonlinearity $\gamma$ is
increased one of the bound states becomes more and more localized,
while the other becomes more and more delocalized. Since, in the
limit of high nonlinearity, the effective coupling among sites is
negligible, one would expect the bound state to become more and
more localized. Therefore, the state with smaller localization
length is stable, the other unstable. This qualitative argument is
confirmed by the more rigorous procedure of examining the
Hamiltonian flow of the system around the two fixed points (bound
states).

Figure 3 shows the critical nonlinearity for the onset of a bound state, as a function of the
distance from the impurity to the lattice `surface' ($n=0$), for different nonlinearity exponents.
Significative differences from the infinite lattice case are apparent: As the impurity is
placed closer and closer to the surface, the critical nonlinearity to create a bound state increases
or decreases, depending upon whether the nonlinear exponent is below or above, approximately two.
In particular, for the all-important standard DNLS case, $\beta = 2$, the presence of a surface increases
the nonlinearity needed to create a bound state, contrary to popular belief that a surface would help
localize the electron.

For a given value of exponent $\beta$ and any inclusion distance
$d$, the bound state probability profile $|C_{n}|^{2}$ is given in
closed form by $|C_{n}|^{2} = A [\ Q^{|n-d|} - Q^{|n+d+2|}\ ]$, 
where $Q\equiv
z_{b}-\sqrt{z_{b}^{2}-1}$, $A\equiv (z_{b}-Q)/(z_{b}+(z_{b} + 2
(1+d)\sqrt{z_{b}^{2}-1}) Q^{2(1+d)})$ and $z_{b}$ is the solution
of Eq.(\ref{eq:zb}). Simple analysis of this profile shows
that $|C_{n}|^{2}$ has always a single hump at $n = d$. This
profile is shown in Fig.4 for the standard DNLS ($\beta = 2$) and
a nonlinearity strength $\gamma$ just above critical,
at four different impurity locations under the surface. its
general features are shared by other $\beta$ exponents. 
\begin{figure}[th]
\centerline{\includegraphics[height=2.in,width=3.in]{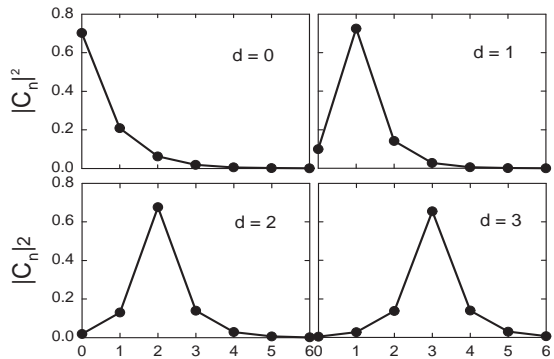}}
\caption{Probability profile for the stable bound state at
different impurity positions ($\beta = 2$, $\gamma = 1.305$).}
\label{fig4}
\end{figure}
Below the surface, the probability profile converges quickly to the infinite
lattice case as $d$ increases past $4$.

{\em Dynamical Seltrapping.}\  We compute numerically the long-time average probability at the impurity site,
$P_{d} = \lim_{T\rightarrow \infty} (1/T) \int_{0}^{T} dt | C_{d} |^{2}$,
for several distances $d$ from the `surface' ($n = 0$). As initial condition,
we use a completely localized excitation on the impurity site: $C_{n}(0) = \delta_{n d}$.
Figure 5 shows the critical
nonlinearity for selftrapping ($P_{d} > 0$) as a function of the
distance between the nonlinear impurity and the chain surface, for different
nonlinearity exponent values. In general, the behavior is qualitatively similar to the one observed
for the onset of a bound state (Fig.3). In both cases, for a fixed distance,
an increase of the nonlinear exponent $\beta$ results
on an increase of the nonlinearity threshold for selftrapping.
The same behavior was observed previously for an impurity in a completely infinite
chain\cite{mth_pre}. This phenomenon is not hard to explain:
Since $|C_{n}| < 1$, we see from Eq.(\ref{eq:1}) that as $\beta > 0$ is increased,
$|C_{d}|^{\beta}$ will necessarily decrease, which implies that a larger
$\gamma$ will be needed to keep the value of the {\em effective} impurity strength
$\gamma |C_{d}|^{\beta}$.  Thus, at a fixed impurity-surface distance, a higher
$\beta$ implies a higher $ \gamma_{c}$.  Another interesting behavior we observe
from figs.5 and 3 is that for a fixed nonlinearity exponent, the critical nonlinearity
depends roughly on whether the exponent is below or above two, approximately: For
$\beta < 2$, an increase in the impurity-surface distance $d$ results on a
decrease of $\gamma_{c}$, while for $\beta > 2$,  an increase in $d$
increases $\gamma_{c}$.  The explanation of this phenomenon seems to rest on the
delicate balance between kinetic and potential energies. If we assume an
electronic bound state $\Psi$ with localization length $\lambda$, then on
normalization grounds we have $|\Psi|^{2}\sim 1/\lambda$. The kinetic energy
content is $\Delta K \sim h^{2}/2 m \lambda^{2}$, while the average
potential energy is in magnitude equal to
$\Delta V = \int d x\ V(x) |\Psi(x)|^{2} = \int d x\ \gamma |\Psi(x)|^{\beta} |\Psi(x)|^{2} \sim
\gamma a /\lambda^{[1+(\beta/2)]}$, where $a$ is of the order of the lattice spacing.
Thus,
\be
\Delta V/\Delta K \sim \gamma\ \lambda^{1- (\beta/2)}.\label{eq:arg}
\ee
On the other hand, as the impurity
is brought closer to the surface, the wavefunction becomes more `compressed' (Fig.4), i.e., 
$\lambda$ decreases as $d$ approaches zero. This implies, 
from Eq.(\ref{eq:arg}) that  for $\beta > 2$, a decrease in $\lambda$ increases 
$\Delta V$ with respect to $\Delta K$, which means that less nonlinearity is needed to maintain a
localized state. On the contrary, if $\beta < 2$, an decrease in $\lambda$ decreases $\Delta V$ with
respect to $\Delta K$ and now, more nonlinearity is needed to maintain the localized state. 

{\em Completely nonlinear lattice.}\  In the large
nonlinearity limit where $\gamma \gg \gamma_{cr}$, the single
nonlinear impurity results should approximate those corresponding
to a whole nonlinear lattice. For the particular case examined in
this work, the ``extended'' problem
\begin{figure}[ht]
\centerline{\includegraphics[height=2.in,width=3.in]{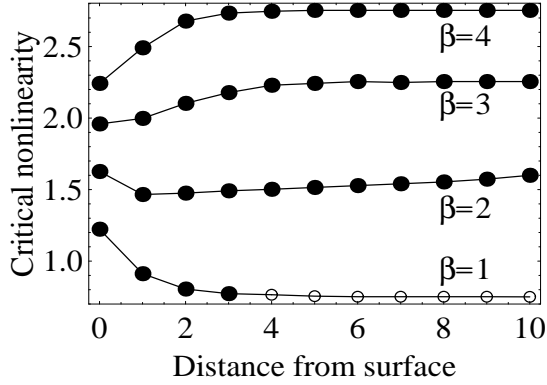}}
\caption{Right:\ Scaled critical nonlinearity for dynamical selftrapping
as a function of the distance from the impurity to the chain
surface, for several nonlinearity exponents. The empty circles
shown for $\beta = 1$ and $d =4$ through $d = 10$ indicate
approximate values since the selftrapping is not abrupt.}
\label{fig5}
\end{figure}
consists of the formation of
an intrinsic localized mode (ILM) in a semi-infinite nonlinear
latice. Due to the presence of a surface, the discrete
translational invariance is broken and a natural question arises:
where will the localized state be formed? Our single nonlinear
impurity analog can provide an answer. For each impurity position
$d$, the bound state energy can be computed as a function of $d$.
The position corresponding to its minimum value will correspond to
the position of the ILM. Also, the impurity energy and spatial
probability profile should approximate the ones corresponding to
the ILM. Figure 6 shows the impurity energies as a function of
distance from the lattice surface, for the DNLS case, $\beta = 2$
and $\gamma = \pm 2$. We see that for a
positive value of the nonlinearity parameter  $\gamma$, the
preferred position is the very surface ($d=0$), while for a
negative $\gamma$, the preferred position is one layer below the
surface ($d=1$). These predictions are indeed confirmed by direct
numerical computations, where the Hamiltonian corresponding to a
semi-infinite nonlinear lattice $H = (1/2) \sum_{n=0}^{\infty} (
|n\rangle \langle n+1| + |n+1\rangle \langle n| ) +
    \gamma \sum_{n=0}^{\infty}|C_{n}|^{\beta} |n\rangle \langle
    n|$, is diagonalized by an iterative procedure. For the particular
example of fig. 6, the error obtained for the ILM energy is about
$1\%$.

\begin{figure}[t]
\centerline{\includegraphics[height=2.in,width=2.5in]{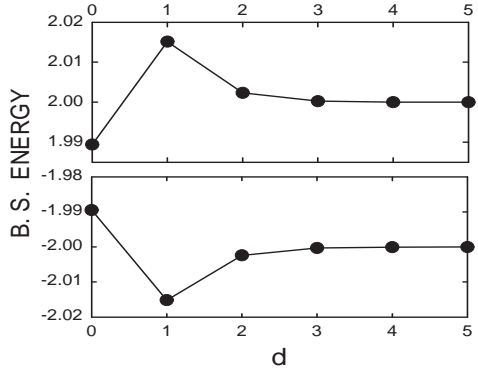}}
\caption{Nonlinear impurity bound state energy as a function of
distance impurity-surface, for $\beta = 2$ and $\gamma = 2$ (upper)
and $\gamma = -2$ (lower).} \label{fig6}
\end{figure}

\section*{Acknowledgements}

This work was partially supported by Fondecyt grant 1020139. The
author is grateful to R. Seiringer for useful discussions.

\end{document}